\documentclass[aps,pra,preprint,amsmath]{revtex4-1}
\usepackage{silence}
\usepackage{mathrsfs}
\usepackage{graphicx}
\usepackage{epstopdf}
\usepackage{amsthm}
\usepackage{amsmath}
\usepackage{amsfonts}
\usepackage{amssymb}
\usepackage{braket}
\usepackage{bbold}
\usepackage{subfigure}
\usepackage{appendix}
\usepackage{tikz}
\usepackage{bm}
\usepackage{float}
\def\be{\begin{equation}}
\def\ee{\end{equation}}
\def\ba{\begin{array}}
\def\ea{\end{array}}

\def\1{{\bf{1}}}

\input amssym.def
\begin{document}
\title{\bf A Note on Lower Bounds of Concurrence for Arbitrary Dimensional Bipartite Quantum States}

\author{Zhi-Bo Chen}
\author{Shao-Ming Fei}

\email{feishm@cun.edu.cn}
\affiliation{
School of Mathematical Sciences, Capital Normal University, Beijing 100048, China
}

\begin{abstract}
Quantification of quantum entanglement plays a crucial role in the study of quantum information tasks. We present analytical lower bounds for both concurrence and 2-concurrence based on the correlation matrices of bipartite quantum states. Compared with other related lower bounds, our approach provides a better estimation of the entanglement, particularly for states with large purity.

\end{abstract}

\maketitle

%\noindent {\bf 1. Introduction}

Quantum entanglement is a remarkable property of quantum systems, serving as fundamental physical resources in quantum information processing, including quantum cryptography, quantum teleportation and quantum computation \cite{QCQI}. The resource theory of quantum entanglement has been well established with the LOCC (local operation and classical communication) \cite{IE} as the free operations and the separable states as the free states. To quantify the entanglement, many entanglement measures have been presented. However, most entanglement measures have no explicit analytical formulas for general quantum mixed states, due to the optimization involved in the definition with convex roof extension.

An important entanglement measure is the concurrence \cite{Intro Concurrence,af}. For a bipartite pure state $\ket{\psi} \in \mathbb{C}^{d_1} \otimes \mathbb{C}^{d_2}$ in systems $A$ and $B$, the concurrence of $\ket{\psi}$ is defined by $C(\ket{\psi}):= \sqrt{2(1-Tr({\rho_A}^2))}$, where $\rho_A = Tr_B \ket{\psi}\bra{\psi}$ is the reduced state of subsystem $A$ obtained by tracing over the subsystem $B$. The concurrence is then extended to mixed states $\rho$ by the convex roof extension,
$$
C(\rho)= \underset{\{ P_i, \ \ket{\psi} \}}{min} \sum_i P_i C(\ket{\psi_i}),
$$
where the minimization goes over all possible pure state decompositions of $\rho = \sum_i P_i \ket{\psi_i}\bra{\psi_i}$ \cite{QCQI}. Generally it is a challenge to calculate exactly the concurrence for general mixed states. Many lower bounds of concurrence have been presented \cite{Concurrence PRL, Concurrence PRA, Concurrence old, Concurrence KF, Concurrence 1, Concurrence QIP, Concurrence 2, Concurrence 3}. Besides, the lower bounds of $q$-concurrence \cite{q-Concurrence} have been also studied, where the $q$-concurrence of a pure state is defined by $C_q(\ket{\psi})=1-Tr({\rho_A}^q)$, $q \geq 2$. The $q$-concurrence of mixed states is defined similarly by the convex roof extension.

%\noindent {\bf 2. Main Results}

In this paper, we present new lower bounds of concurrence and 2-concurrence based on the correlation matrices of bipartite quantum states. For an arbitrary bipartite quantum state $\rho\in\mathbb{C}^{d_1} \otimes \mathbb{C}^{d_2}$, $\rho$ has a generalized Bloch representation \cite{KF1,KF2},
\begin{align}\label{p1}
\rho = \frac{1}{d_1 d_2} \left (  I_{d_1 d_2} + \sum_{i,j \ \mbox{\scriptsize{are not all 0} } } u_{ij} \ \widetilde{\lambda_i}^{(1)} \otimes \widetilde{\lambda_j}^{(2)}  \right ),
\end{align}
where $\widetilde{\lambda_0}^{(1)}= \sqrt{d_1 -1} I_{d_1}$, $\widetilde{\lambda_0}^{(2)}= \sqrt{d_2 -1} I_{d_2}$, $\widetilde{\lambda_i}^{(1)}= \sqrt{\frac{d_1(d_1 -1)}{2} } \lambda_i^{(1)} $,  $\widetilde{\lambda_j}^{(2)}= \sqrt{\frac{d_2(d_2 -1)}{2} } \lambda_j^{(2)} $, $\left\{ \lambda_i^{(1)} \right\}_{1 \leq i \leq {d_1}^2 -1}$ and $\left\{ \lambda_j^{(2)} \right\}_{1 \leq i \leq {d_2}^2 -1}$ are generators of $SU(d_1)$ and $SU(d_2)$, respectively, satisfying $Tr(\lambda_i \lambda_j) = 2\delta_{ij}$ and $Tr(\lambda_i) = 0$, $u_{ij}  = \frac{1}{(d_1 -1)(d_2 -1)} Tr \left( \rho \cdot \widetilde{\lambda_i}^{(1)} \otimes \widetilde{\lambda_j}^{(2)} \right) $. Eq.(\ref{p1}) can be further rewritten as
\begin{align}\label{p2}
\rho = \frac{1}{d_1 d_2} \left (  I_{d_1 d_2} + \sum_{i=1}^{{d_1}^2-1} r_i \ \widetilde{\lambda_i}^{(1)} \otimes \widetilde{\lambda_0}^{(2)} +
\sum_{j=1}^{{d_2}^2-1} s_i \ \widetilde{\lambda_0}^{(1)} \otimes \widetilde{\lambda_j}^{(2)} +
\sum_{i,j \neq 0 } t_{ij} \ \widetilde{\lambda_i}^{(1)} \otimes \widetilde{\lambda_j}^{(2)}   \right ).
\end{align}
Denote $R = (r_i)$ and $S = (s_j)$ the real vectors, and $T = (t_{ij})$ the $({d_1}^2-1) \times ({d_2}^2-1)$ correlation matrix. Set $\boldsymbol{u} = (u_{ij})$ a real vector, called the vector representation of $\rho$. Then $\boldsymbol{u}=(R,S,vec(T))$ is the vector representation of $\rho$, where $vec(T)$ is vectorization of $T$.

For a matrix $A=a_{ij}$, the Frobenius norm $\left\| A \right\|_F= \sqrt{Tr(A A^\dagger)}$. It is verified that ${\left\| A \right\|_F}^2 = \sum_{i,j} |a_{ij}|^2 = |vec(A)|^2$. For the vector representation $\boldsymbol{u} = (R,S,vec(T))$, we have
\begin{align}\label{p3}
|\boldsymbol{u}|^2 = \sum_{i,j \ \mbox{\scriptsize{are not all 0} } } {u_{ij}}^2 = |R|^2 + |S|^2 + {\left\| T \right\|_F}^2 .
\end{align}
Furthermore, it is straightforward to verify that $\left\| T \right\|_F = \sqrt{\left\| T T^\dagger \right\|_{tr}}$, where $\left\| A \right\|_{tr} = Tr( \sqrt{A A^\dagger} )$ is trace norm, also known as Ky-Fan norm. It is easily verified that $\left\| T \right\|_F \leq \left\| T \right\|_{tr}$. Based on $\left\| T \right\|_{tr}$, necessary separability conditions and lower bounds of concurrence \cite{Concurrence KF} have been derived. A quantum state is entangled when $\left\| T \right\|_{tr}$ is sufficiently large \cite{KF1,KF2,KF3,KF4}.

Interestingly, the norm $\left\| T \right\|_{F}$ can also gives rise to lower bounds of concurrence. We first prove that the 2-norm of the vector representation of $\rho$ and the purity of $\rho$ are equivalent.

{\bf Lemma 1 \ }
For bipartite state $\rho\in\mathbb{C}^{d_1} \otimes \mathbb{C}^{d_2}$ given by (\ref{p2}), we have
$$
\begin{array}{rcl}
Tr(\rho^2) &=& \displaystyle\frac{1}{d_1 d_2} \left ( 1 + (d_1-1) (d_2 -1) |\boldsymbol{u}|^2 \right ),\\[3mm]
Tr(\rho_A^2) &=& \displaystyle\frac{1}{d_1} \left ( 1 + (d_1-1) (d_2 -1) \cdot |R|^2 \right ),\\[3mm]
Tr(\rho_B^2) &=& \displaystyle\frac{1}{d_2} \left ( 1 + (d_1-1) (d_2 -1) \cdot |S|^2  \right ). \end{array}
$$

{\it Proof \ }
By (\ref{p1}) we have
\begin{align} \notag
Tr(\rho^2)  &= \frac{1}{(d_1 d_2)^2} \left (  d_1 d_2 + d_1 d_2 (d_1 -1) (d_2 -1) \sum_{i,j} \sum_{k,l} u_{ij} u_{kl} \delta_{ik} \delta_{jl}  \right )  \notag \\
& = \frac{1}{(d_1 d_2)^2} \left (  d_1 d_2 + d_1 d_2 (d_1 -1) (d_2 -1) \sum_{i,j} {u_{ij}}^2  \right )  \notag \\
& =  \frac{1}{d_1 d_2} \left (  1+ (d_1 -1) (d_2 -1) |\boldsymbol{u}|^2  \right ).\notag
\end{align}
By (\ref{p2}) we have
$$ \rho_A = \frac{1}{d_1} \left (  I_{d_1} + \sqrt{d_2 -1} \sum_{i=1}^{{d_1}^2-1} r_i \ \widetilde{\lambda_i}^{(1)}  \right ) , $$
then
\begin{align} \notag
Tr( {\rho_A}^2 ) & = \frac{1}{{d_1}^2} \left ( Tr ( {I_{d_1}}^2 ) + (d_2 -1) \sum_{i=1}^{{d_1}^2-1} \sum_{j=1}^{{d_1}^2-1} r_i r_j Tr\left ( \widetilde{\lambda_i}^{(1)} \widetilde{\lambda_j}^{(1)} \right ) +2 \sum_{i=1}^{{d_1}^2-1} Tr( \widetilde{\lambda_i}^{(1)} )  \right )  \\
& = \frac{1}{d_1} \left ( 1 + (d_1 -1)(d_2 -1) \sum_{i=1}^{{d_1}^2-1} {r_i}^2 \right ) \notag \\
&= \frac{1}{d_1} \left ( 1 + (d_1 -1)(d_2 -1) \cdot |S|^2 \right ).\notag
\end{align}
$Tr( {\rho_B}^2 )$ can be similarly derived.  $\Box$

Based on the generalized Bloch representation, the concurrence of a pure state is given by the Frobenius norm of the correlation matrix.

{\bf Theorem 1 \ }
For any bipartite pure state $\ket{\psi} \in \mathbb{C}^{d_1} \otimes \mathbb{C}^{d_2}$, we have
\begin{align} \notag
C(\ket{\psi}) &= \sqrt{ \frac{2(d_1 -1)(d_2 -1)}{d_1+d_2} \left (  {\left\| T \right\|_{F}}^2 -1 \right ) },\\
C_2(\ket{\psi}) &=  \frac{(d_1 -1)(d_2 -1)}{d_1+d_2} \left (  {\left\| T \right\|_{F}}^2 -1 \right ),\notag
\end{align}
where $T$ is correlation matrix of $\ket{\psi}\bra{\psi}$.

{\it Proof \ }
For a pure state $\rho=\ket{\psi}\bra{\psi}$, from Lemma 1 and Eq.(\ref{p3}) we have
\begin{align}\label{p4}
|R|^2 + |S|^2 + {\left\| T \right\|_{F}}^2 = \frac{d_1 d_2 -1}{(d_1 -1)(d_2 -1)}.
\end{align}
Since $Tr({\rho_A}^2) = Tr({\rho_B}^2)$ for pure state \cite{QCQI}, by Lemma 1 we have
$$ |R|^2 = \frac{d_1 Tr({\rho_A}^2) -1}{(d_1 -1)(d_2 -1)},~ \      |S|^2 = \frac{d_2 Tr({\rho_A}^2) -1}{(d_1 -1)(d_2 -1)}.$$
From Eq.(\ref{p4}) we have
\begin{align}
{\left\| T \right\|_{F}}^2 - 1 & = \frac{d_1 d_2 -1}{(d_1 -1)(d_2 -1)} - |R|^2 - |S|^2 -1  \notag \\
& = \frac{ d_1d_2 -1 - (d_1Tr({\rho_A}^2) -1) - (d_2Tr({\rho_A}^2) -1) -(d_1 -1)(d_2 -1) }{(d_1 -1)(d_2 -1)}   \notag \\
& = \frac{d_1 + d_2}{(d_1 -1)(d_2 -1)} \left ( 1 -Tr({\rho_A}^2) \right ) .   \notag
\end{align}
Therefore, $1 -Tr({\rho_A}^2) = \frac{(d_1 -1)(d_2 -1)}{d_1 + d_2} \left ( {\left\| T \right\|_{F}}^2 - 1 \right )$, which proves the theorem.  $\Box$

For general bipartite mixed states, we have the following conclusion.

{\bf Theorem 2 \ }
For any bipartite state $\rho\in\mathbb{C}^{d_1} \otimes \mathbb{C}^{d_2}$, we have
$$
C(\rho)  \geq  \frac{\sqrt{2}(\sqrt{K +1} +1)}{K} \left ( \left\| T \right\|_{F} -1 \right ),
$$
$$
C_2(\rho) \geq  \frac{(d_1 -1)(d_2 -1)}{d_1+d_2} \left (  {\left\| T \right\|_{F}}^2 -1 \right),
$$
where $T$ is the correlation matrix of $\rho$, $K = \frac{d_1+d_2}{(d_1 -1)(d_2 -1)}$.

{\it Proof \ }
Since $\sqrt{2} > C(\ket{\psi}) \geq 0$, by Theorem 1 we have $2 > \frac{2(d_1 -1)(d_2 -1)}{d_1+d_2} \left (  {\left\| T \right\|_{F}}^2 -1 \right ) \geq 0$. Hence, the correlation matrix of a pure state always satisfies
\begin{align}\label{p5}
\sqrt{ \frac{d_1+d_2}{(d_1 -1)(d_2 -1)} +1}  >  \left\| T \right\|_{F}  \geq 1.
\end{align}
It is easy to verify that if $a \geq x \geq 1$, then $\sqrt{x^2 -1} \geq \sqrt{\frac{a+1}{a-1}} (x-1)$. Take $a = \sqrt{K +1}$, where $K = \frac{d_1+d_2}{(d_1 -1)(d_2 -1)}$, by (\ref{p5}) we have
\begin{align} \notag
\sqrt{ { \left\| T \right\|_{F} }^2 -1 } \geq \sqrt{\frac{\sqrt{K +1} +1}{\sqrt{K +1} -1}} ( \left\| T \right\|_{F} -1 ).
\end{align}
From Theorem 1 we have
\begin{align}\label{p6}
C(\ket{\psi}) = \sqrt{ \frac{2}{K} \left (  {\left\| T \right\|_{F}}^2 -1 \right ) }  \geq  \sqrt{ \frac{2}{K} \cdot \frac{\sqrt{K +1} +1}{\sqrt{K +1} -1}} ( \left\| T \right\|_{F} -1 ).
\end{align}

Next consider the optimal decomposition $\rho = \sum_i P_i \ket{\varphi_i}\bra{\varphi_i}$ for the concurrence. Using (\ref{p6}) we have
\begin{align} \notag
C(\rho) = \sum_i P_i C(\ket{\varphi_i}) & \geq \sqrt{ \frac{2(\sqrt{K +1} +1)}{K(\sqrt{K +1} -1)}} \sum_i P_i \left ( \left\| T_i \right\|_{F} -1 \right ) \\
& \geq  \sqrt{ \frac{2(\sqrt{K +1} +1)}{K(\sqrt{K +1} -1)}} \left ( \left\| T \right\|_{F} -1 \right ) =  \frac{\sqrt{2}(\sqrt{K +1} +1)}{K} \left ( \left\| T \right\|_{F} -1 \right ) ,
 \notag
\end{align}
where $T_i$ is correlation matrix of the state $\ket{\varphi_i}\bra{\varphi_i}$.

For the second inequality in Theorem 2, consider the optimal decomposition $\rho = \sum_i P_i \ket{\psi_i}\bra{\psi_i}$ for the 2-concurrence. Using Theorem 1 we have
\begin{align} \notag
C_2(\rho) = \sum_i P_i C_2(\ket{\psi_i}) = \frac{(d_1 -1)(d_2 -1)}{d_1+d_2} \sum_i P_i \left (  {\left\| T_i \right\|_{F}}^2 -1 \right ),
\end{align}
where $T_i$ is correlation matrix of the state $\ket{\psi_i}\bra{\psi_i}$. Since $f(A) := {\left\| A \right\|_{F}}^2$ is convex function, we have
$$
C_2(\rho) \geq  \frac{(d_1 -1)(d_2 -1)}{d_1+d_2}  \left (  {\left\| \sum_i P_i T_i \right\|_{F}}^2 -1 \right )  =  \frac{(d_1 -1)(d_2 -1)}{d_1+d_2}  \left (  {\left\| T \right\|_{F}}^2 -1 \right ).
$$
$\Box$

The lower bound of the 2-concurrence in Theorem 2 is exact for pure states. One would expect that Theorem 2 gives a better lower bound for states of large purity. Interestingly, this is also true for the lower bound of concurrence in Theorem 2.

{\bf Example 1 } Consider the $4 \times 4$ state $\rho_x = x \ket{\phi}\bra{\phi} + \frac{1-x}{16} I_{16}$, where $\ket{\phi} = \frac{1}{\sqrt{2}} \left ( \ket{00}+\ket{33} \right )$ and $0<x<1$. For this state we have ${\left\| T \right\|_{F}}^2 = \frac{13}{9} x^2$. By Theorem 2 we obtain
$$C(\rho_x) \geq \frac{9 \sqrt{2}}{8} \left ( \frac{\sqrt{17}}{3} +1 \right ) \left ( \frac{\sqrt{13}}{3}x -1 \right ),~~~~~ C_2(\rho_x) \geq \frac{9}{8} \left ( \frac{13}{9}x^2 -1 \right ).$$
In \cite{Concurrence PRL} a lower bound of concurrence has been presented for $m \times n$ ($m \leq n$) states,
$$
C(\rho) \geq \sqrt{\frac{2}{m(m-1)}} \left ( max \left\{  \left\| \rho^{T_A} \right\|_{tr}, \ \left\| \mathcal{R}(\rho) \right\|_{tr}  \right\} -1 \right ) \ ,
$$
where $T_A$ is partial transposition with respect to the first subsystem \cite{PPT}, $\mathcal{R}$ is realignment operation \cite{R}. For $\rho = \sum_{ijkl} \rho_{ij,kl} \ket{ij}\bra{kl}$, $\rho^{T_A} = \sum_{ijkl} \rho_{ij,kl} \ket{kj}\bra{il}$ and $\mathcal{R}(\rho) = \sum_{ijkl} \rho_{ij,kl} \ket{ik}\bra{jl}$. For the state $\rho_x$, one has $C(\rho_x) \geq \frac{1}{\sqrt{6}} \left ( \frac{7+9x}{8} -1 \right )$. When $x > 0.914$, the lower bound of $C(\rho_x)$ given in Theorem 2 is significantly better than the one given in \cite{Concurrence PRL}, as shown in Fig.1.
\begin{figure}[!hbtp]
    \centering
    \includegraphics[width=0.5\linewidth]{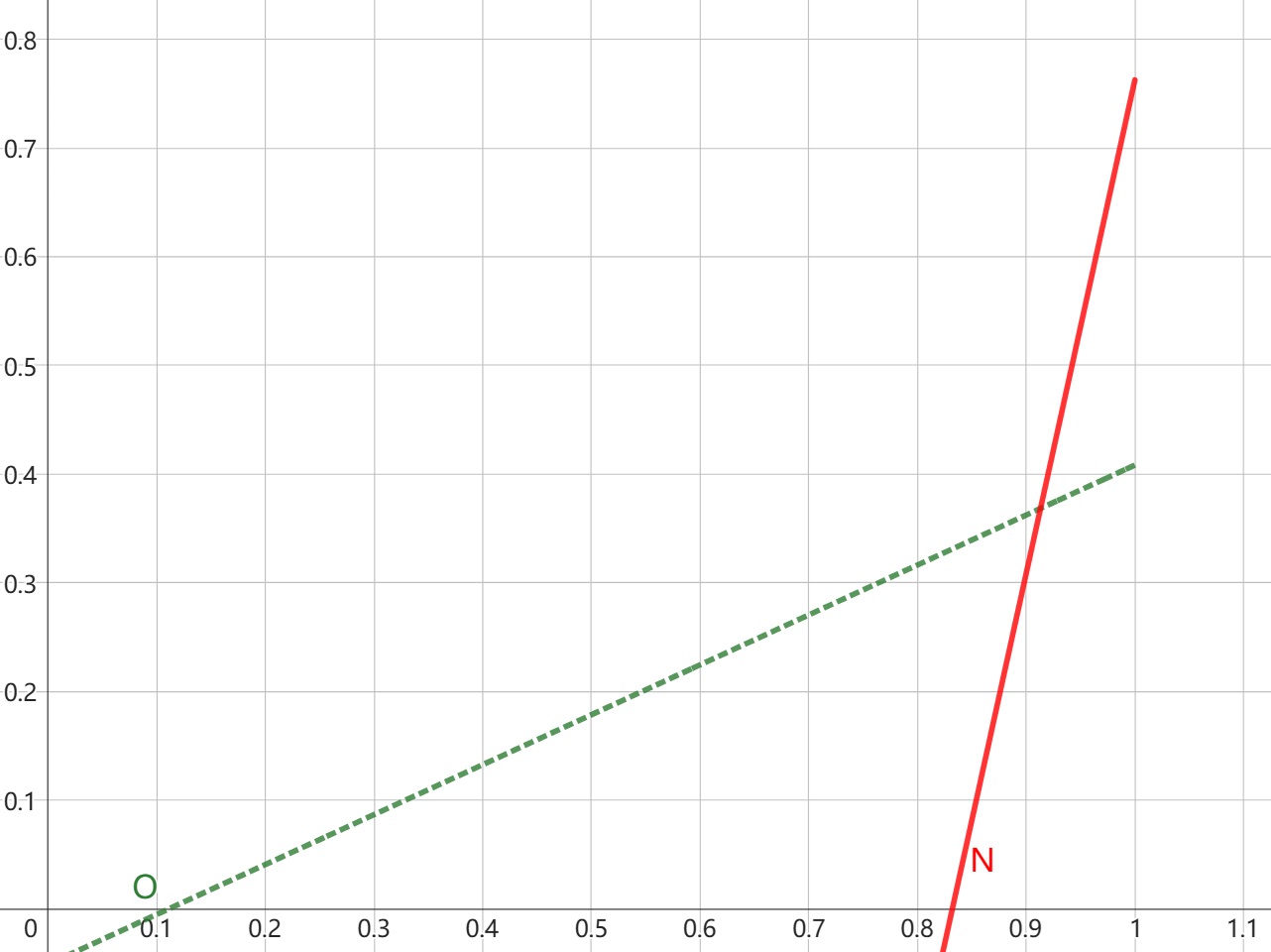}
    \caption{Red solid line (N) is for the lower bound of concurrence from Theorem 2 of this paper, green dashed line (O) is for that from \cite{Concurrence PRL}.}
    \label{fig:enter-label}
\end{figure}

In \cite{q-Concurrence} the authors presented a lower bound of 2-concurrence for $m \times n$ ($m \leq n$) states,
$$
C_2(\rho) \geq \frac{1}{m(m-1)} \left ( max \left\{  \left\| \rho^{T_A} \right\|_{tr}, \ \left\| \mathcal{R}(\rho) \right\|_{tr}  \right\} -1 \right )^2.
$$
For the state $\rho_x$, one has $C_2(\rho_x) \geq \frac{1}{12} \left ( \frac{7+9x}{8} -1 \right )^2$. When $x > 0.854$, the lower bound of $C_2(\rho_x)$ given in Theorem 2 is significantly better than the one given in \cite{q-Concurrence}, see Fig.2.
\begin{figure}[!hbtp]
    \centering
    \includegraphics[width=0.5\linewidth]{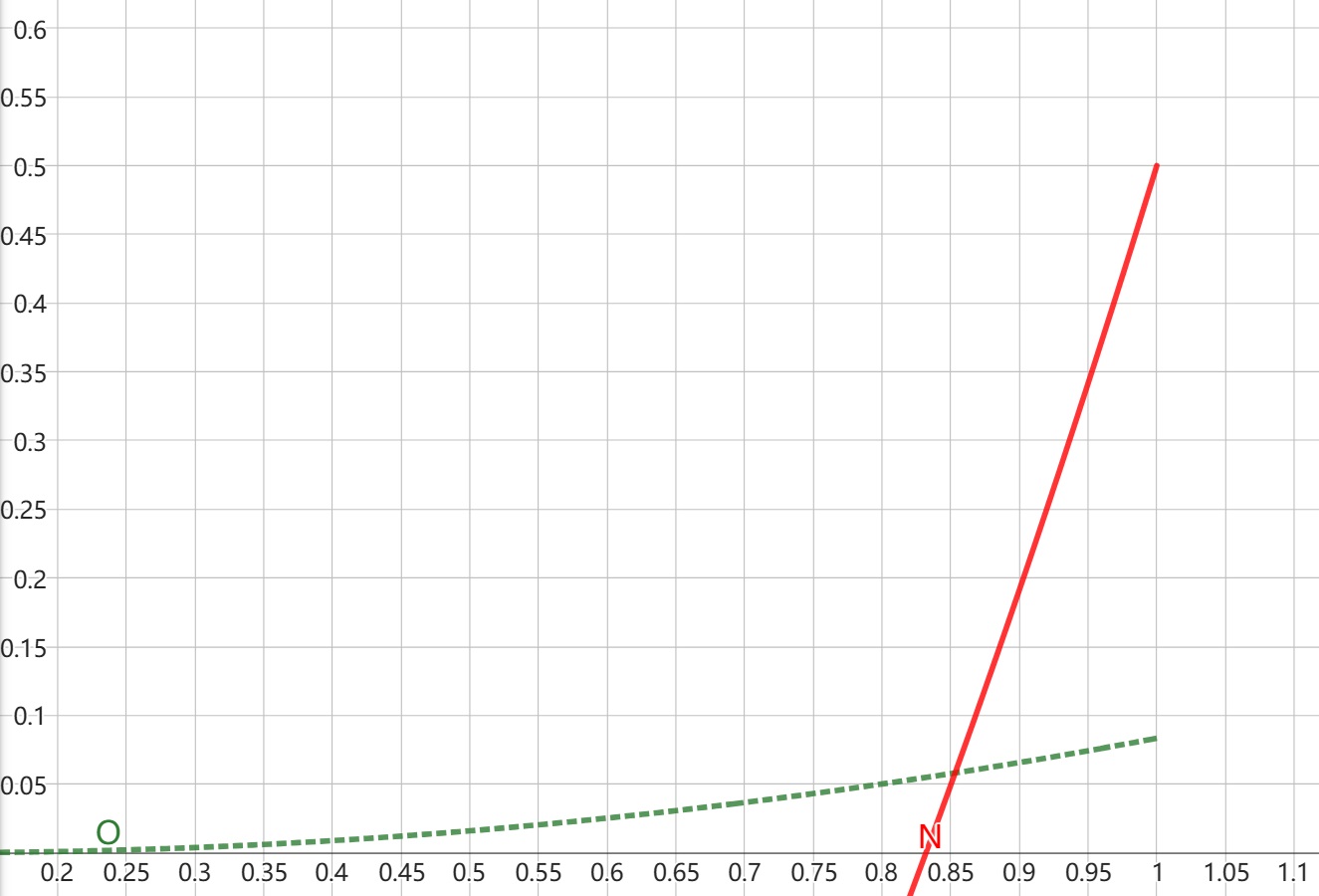}
    \caption{Red solid line (N) is for the lower bound of 2-concurrence from Theorem 2 of this paper, green dashed line (O) is for that from the Theorem 1 in \cite{q-Concurrence}.}
    \label{fig:enter-label}
\end{figure}

Example 1 seems to imply that Theorem 2 is not a good lower bound of concurrence for states with smaller purity. But for states of large purity, Theorem 2 gives a better lower bound of concurrence. Theorem 2 gives a lower bound of concurrence only when $\left\| T \right\|_{F} > 1$, it is complementary to the lower bound presented in \cite{Concurrence PRL,q-Concurrence}. As Fig.1 and Fig.2 show, the two curves together give a better lower bound of concurrence.

{\bf Example 2 }
Consider the following state given in \cite{Concurrence PRA},
$$
\rho = \frac{1}{4} diag  \left ( q_1, q_4, q_3, q_2, q_2, q_1, q_4, q_3, q_3, q_2, q_1, q_4, q_4, q_3, q_2, q_1 \right ) + \frac{q_1}{4} \sum_{i,j=1,6,11,16}^{i\neq j} E_{ij} \ ,
$$
where $E_{ij}$ is a matrix with the $(i,j)$ entry 1 and the rest of the entries $0$, $q_m \geq 0$, and $\sum_{m=1}^4 q_m = 1$. Take $q_2 = q_4 = \frac{1}{2} (1-q_1), \ q_3=0$. We have ${\left\| T \right\|_{F}}^2 = \frac{1}{9} \left ( 18 {q_1}^2 -4 q_1 +1 \right )$. From Theorem 2, we have the lower bound of concurrence,
$$ C(\rho) \geq \frac{9\sqrt{2}}{8} \left ( \sqrt{\frac{17}{9}} +1 \right )  \left ( \frac{1}{3} \sqrt{18 {q_1}^2 -4 q_1 +1} -1 \right ). $$
The lower bound from (31) of Ref.\cite{Concurrence PRA} is given by $ C(\rho) \geq \frac{1}{2\sqrt{6}} \left ( \frac{2}{3} q_1 -\frac{1}{6} + |\frac{2}{3} q_1 -\frac{1}{6}| \right ) $. From \cite{Concurrence old} the lower bound is given by $ C(\rho) \geq \frac{1}{4\sqrt{6}} \left ( \frac{1}{2} ( 3q_1 -1) + |\frac{1}{2} ( 3q_1 -1)| \right ) $. When $q_1 >0.82$, our lower bound of concurrence is significantly better than all of them, as shown in Figure.3.
\begin{figure}[!hbtp]
    \centering
    \includegraphics[width=0.5\linewidth]{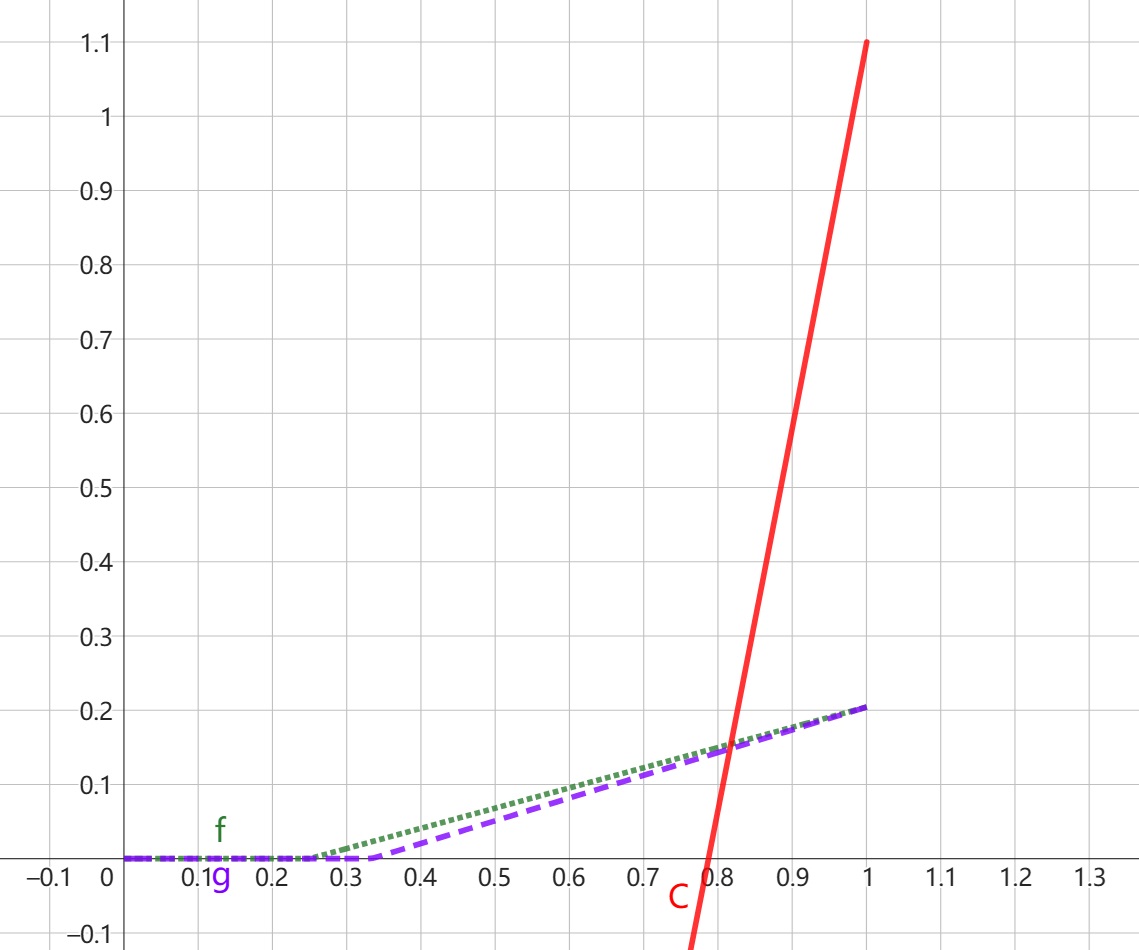}
    \caption{Red solid line (C) is the lower bound of concurrence from Theorem 2 of this paper; green dashed line (f) is the lower bound of concurrence from \cite{Concurrence PRA}; purple dashed line (g) is the lower bound of concurrence from \cite{Concurrence old}.}
    \label{fig:enter-label}
\end{figure}

The examples show that our lower bound of concurrence is significantly larger than other lower bounds when $\left\| T \right\|_{F} > 1$. Our lower bound of concurrence may not be good at entanglement detection, but it gives a better entanglement estimation for quantum states that are more entangled. Such lower bounds of entanglement give another kind of estimation of entanglement.

\medskip

%=============================================================================%
\noindent{\bf Acknowledgements} This work is supported by the National Natural Science Foundation of China under Grants 12075159 and 12171044, the specific research fund of the Innovation Platform for Academicians of Hainan Province.

%===========================================================================%

%=============================================================================%

%===========================================================================%

%=============================================================================%
%===========================================================================%

\end{document}